\documentclass{rmaa}

\usepackage{paralist}
\usepackage{psfrag,color}

\title{The interaction of a jet-like pulse with a wind from a stellar
companion} 

\author{
  A. Riera, \altaffilmark{1,2}
  A. C. Raga \altaffilmark{3} 
  J. Alcolea \altaffilmark{4}
  \medskip}

\altaffiltext{1}{Departament de F\'\i sica i Enginyeria
Nuclear, Universitat Polit\`ecnica de Catalunya,
Spain.}
\altaffiltext{2}{Departament d'Astronomia i Meteorologia. Universitat 
de Barcelona, Spain}
\altaffiltext{3}{Instituto de Ciencias Nucleares, Univer\-si\-dad
Nacio\-nal Aut\'o\-noma de M\'e\-xico, M\'exico.}
\altaffiltext{4}{Observatorio Astron\'omico Nacional (OAN), Apdo. 1143, 28800 
Alcal\'a de Henares, Spain}

\shortauthor{Riera, Raga \& Alcolea}
\shorttitle{Jet/wind interaction}

\fulladdresses{
\item Javier Alcolea: Observatorio Astron\'omico Nacional, Apdo. 1143, 28800 
Alcal\'a de Henares, Spain.
  (j.alcolea@oan.es). 
\item Alejandro C. Raga: Instituto de Ciencias Nucleares, UNAM,
 Apdo. Postal 70-543, 04510 M\'exico, D. F., M\'exico. 
  (raga@nuclecu.unam.mx)
\item Angels Riera: Departament de F\'\i sica i Enginyeria
Nuclear, Universitat Polit\`ecnica de Catalunya, Av. V\'\i ctor
Balaguer s/n, E-08800 Vilanova i la Geltr\'u, Spain.
  (angels.riera@upc.es).
 }

\listofauthors{J. Alcolea, A.C. Raga, A. Riera}
\indexauthor{Alcolea, J.}
\indexauthor{Raga, A. C.}
\indexauthor{Riera, A.}

\ReceivedDate{} \AcceptedDate{}

\abstract{We present numerical simulations of the interaction 
between a collimated, bipolar ``pulse'' ejected from a star 
and a continuous wind ejected  from a stellar companion. 
We explore the characteristics of the predicted H$\alpha$ intensity maps  
by varying selected input parameters. We find that the asymmetry (in size and 
strength) between the two lobes of the proto-planetary nebula 
OH 231.8+4.2 is reproduced in this scenario if the wind ejected by the companion 
star has a strong latitude dependence.}

\resumen{Presentamos simulaciones num\'ericas de la interacci\'on 
entre un ``pulso'' bipolar colimado expulsado por una estrella y el 
viento continuo debido a una estrella compa\~nera. Se muestra 
la variaci\'on de las propiedades de los mapas de intensidad en H$\alpha$ 
al variar los valores de algunos de los par\'ametros de entrada. 
Mostramos que la asimetr\'\i a (en el tama\~no y la intensidad) entre los dos l\'obulos 
de la nebulosa proto-planetaria OH 231.8+4.2 se puede explicar en el escenario propuesto si 
el viento de la estrella compa\~nera muestra una fuerte dependencia con la latitud. }

\addkeyword{ISM: kinematics and dynamics - PLANETARY NEBULA: individual 
(OH231.8+4.2)}

\RescaleTitleLengths{0.9}
\begin{document}

\maketitle

\section{Introduction}
Most proto-planetary Nebulae (PPNe) show bipolar
or mutipolar lobes, highly collimated outflows or jets and other complex
structures.  A particularly well studied PPN candidate is the 
nebula associated with the OH 231.8+4.2 source. This is a remarkable example 
of a bipolar PPNe with collimated outflows. 
In H$\alpha$ and forbidden emission line maps, this object presents
a clear axial 
symmetry with the existence of shocked material in a wide, bipolar bubble 
with an obscuring ridge in between the two lobes 
 (Bujarrabal et al. 2002).  Long-slit spectra of this object
show the presence of
 Herbig-Haro like knots at the end of the bipolar lobes, which probably
 trace the interaction between the fast, collimated molecular jet 
and the ambient material (S\'anchez Contreras et al. 2000a).
Contrary to what happens in most PPNe, the two optical lobes have different 
sizes; the Southern lobe is weaker but about two times more extended than the 
Northern lobe (Reipurth 1987; S\'anchez Contreras et al. 2000a). 

In the optical and NIR continuum, the nebula is narrow and elongated 
along its symmetry axis (Alcolea et al. 2001; Meakin et al. 2003). 
The same narrow component is observed in CO and other molecular emission
lines (Alcolea et al. 
2001; S\'anchez Contreras, Bujarrabal \& Alcolea 1997; S\'anchez Contreras et al. 2000b). 
The kinematical properties of this molecular jet are
remarkable; its velocity increases linearly
with distance from the star following a Hubble-like velocity law, 
which is consistent  with a sudden acceleration event 
(Alcolea et al. 2001; Alcolea 2004). 
The Southern molecular jet (as seen in CO emission) is clearly 
more extended than the Northern jet and shows the highest flow velocity.  
 The deprojected velocities show values of up to  430 km s$^{-1}$ in the Southern jet, and up 
to 210 km s$^{-1}$ in the Northern jet   (Alcolea et al. 2001). 
The length along the axis of the OH 231.8+4.2 nebula is $\sim$50$''$, 
which at a distance of 1.5 kpc (Kastner et al. 1992) implies a 
length $>10^{18}$ cm.

The central source of OH 231.8+4.2 is not detected at optical wavelengths because of 
the dusty envelope, but the reflected light has a M9 III spectral type,
with a blue excess that suggests the presence of a warmer companion 
(or a peculiar photospheric structure and/or an accretion disk; Cohen 1981). 
This star is a typical Mira variable at the end of the Asymptotic Giant Branch (AGB)
identified with QX Pup 
(Kastner et al. 1998). Recently, OH 231.8+4.2 has been mapped at high spatial 
resolution in SiO and H$_2$O maser emission
(S\'anchez Contreras et al. 2002; 
G\'omez \& Rodr\'\i guez 2001). 
The position of the SiO and H$_2$O  masers (which are tracing the position of QX Pup), 
 are clearly offset from the axis of the bipolar outflow
(as first pointed out by G\'omez \& Rodr\'\i guez 2001).

The QX Pup star shows the variability and maser emission properties typical of a 
Mira star (or of an OH/IR star)  at the AGB. 
The contradiction between the apparent evolutionary stage of the nebula 
(i.e., PPN) and that of the QX Pup (AGB) star  
has been discussed by many authors (see, e.g., Alcolea et al. 2001).  
Based on the large displacement ($>1000$ AU) between the position of the SiO maser 
and the bipolar axis of the nebula, Alcolea (2004) 
 has proposed that QX Pup may not be directly related to the bipolar nebula. 
As first pointed out by Alcolea (2004), if the QX Pup star were
located in the 
Northern lobe the observed differences between the two lobes possibly
could be explained as a result of the interaction between the northern lobe
and the wind from QX Pup.

In this paper, we present 3D numerical simulations of the interaction 
between a collimated, bipolar ``pulse'' ejected from a star and a continous wind 
ejected from a stellar companion. These simulations are a  
first attempt to reproduce the asymmetry between the two lobes of the 
bipolar PPN OH 238.1+4.2 in terms of a bipolar outflow/stellar wind
interaction model. From the numerical simulations, we obtain predictions of 
H$\alpha$ maps that can be compared with the optical images of OH 238.1+4.2 
(S\'anchez Contreras et al. 2000a; Bujarrabal et al. 2002).  

The paper is organized as follows. In \S~2 we describe the parameters and 
the numerical method which have been used. In  \S~3 we present the results of 
the numerical simulations. In \S~4 we compare our simulations with the observations 
of the PPN OH 231.8+4.2. 

\section{The numerical models}

\subsection{General considerations}
We compute a series of five models of the interaction between
a collimated, bipolar ``pulse'' ejected from a star and a continuous
wind ejected from a stellar companion. The models are computed
in a 3D cartesian coordinate system, with its origin centered in
the middle of the computational domain. The domain has a physical
extent of $1.5\times 10^{18}$~cm along the $z$-axis, and
of $7.5\times 10^{17}$~cm along the $x$- and $y$-axes. Outflow
conditions are applied in all of the domain boundaries.

The computations are carried out with the ``yguaz\'u-a'' code,
which integrates the gasdynamic equations in a binary, adaptive
computational grid. This code is described in detail by
Raga, Villagr\'an-Muniz \& Navarro-Gonz\'alez (2000). In the
version of the code that we have employed, a single rate equation
for neutral hydrogen is integrated (together with the 3D gasdynamic
equations), and a simplified cooling function is calculated as a function
of the density, ionization fraction and temperature (as described
by Raga et al. 2000). This cooling function is then added as a sink
term in the energy equation.

\begin{figure}[!t]
  \includegraphics[width=\columnwidth]{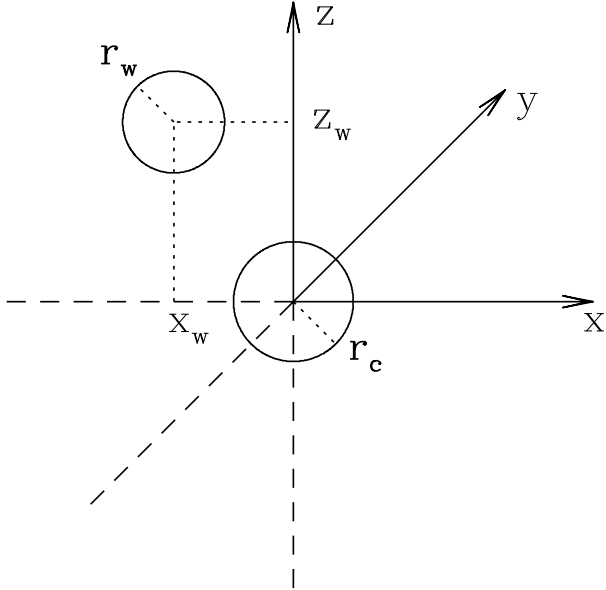}
  \caption{Schematic diagram showing the configuration used
for the numerical simulations (see the text). A collimated wind ``pulse''
is initialized within a sphere of radius $r_c$ (centered at the origin
of the coordinate system) and the continuous
wind from the stellar companion is imposed (at all times) within a sphere
of radius $r_w$, which is centered at the point $(x_w,0,z_w)$. In the
diagram, the projections of these two spheres on the $xz$-plane are
shown. The axis of the ``pulse'' (as well as the axis of the wind in the
simulations with a latitude-dependent wind) is parallel to the $z$-axis.}
  \label{fig1}
\end{figure}

For our simulations, we have used a 5-level binary grid with
a maximum resolution of $5.86\times 10^{15}$~cm along the three axes.
The domain was initialized with a wind solution
(\S 2.2) and a ``bipolar pulse'' (\S 2.3), and the simulations
were carried out until the shell pushed out by the ``pulse'' reaches one
of the boundaries of the computational domain. The initial flow
configuration is shown in Figure \ref{fig1}.

\subsection{The wind}

We have initialized the computational domain with a wind with
a latitude-dependent density of the parametrized form~:
\begin{equation}
\rho={{\dot M}_w\over 4\pi r^2}\,f(\theta)\,,
\label{rho}
\end{equation}
where $r$ is the spherical radius measured from the position of
the wind source. The wind has
an outward directed velocity (independent of $r$) given by~:
\begin{equation}
v={v_w\over f(\theta)}\,,
\label{v}
\end{equation}
with $v_w=40$~km~s$^{-1}$ for all of the computed models, and with
different values of ${\dot M}_w$ for the successive models (see
Table 1). The anysotropy function $f(\theta)$ has the parametrized
form~:
\begin{equation}
f(\theta)=\xi-(\xi-1)\cos^p \theta\,,
\label{f}
\end{equation}
where $\theta$ is the angle measured from the polar axis of the
wind (which we take to be parallel to the $z$-axis of the computational
domain), $\xi$ is a constant which is equal to the equator-to-pole
density ratio, and $p$ is a constant which determines the degree
of flattening towards the equator of the density stratification.
As can be seen from Equation (\ref{f}), a wind with $\xi=1$ is
isotropic. The choices of $\xi$ and $p$ for the computed models
are given in Table 1.

\begin{table}[!t]\centering
  \newcommand{\DS}{\hspace{6\tabcolsep}} 
  \setlength{\tabnotewidth}{\linewidth}
  \setlength{\tabcolsep}{0.7\tabcolsep}
  \tablecols{7}
  \caption{Collimated pulse/wind interaction models}
  \begin{tabular}{ccccccc}
    \toprule
Model & ${\dot M}_w$\tabnotemark{a} & $x_w$\tabnotemark{b}
& $z_w$\tabnotemark{b} & $\xi$\tabnotemark{c} & $p$\tabnotemark{c} &
$t_f$\tabnotemark{d} \\
\phantom{0} & [$10^{-6}$M$_\odot$yr$^{-1}$] &
\multispan2{\hfil [$10^{17}$cm] \hfil} & & & [yr]\\
\midrule
M1 & \phantom{0}1 & $-1.0$ & 1.5 & \phantom{0}1 & --- & 1100 \\
M2 & \phantom{0}5 & $-1.0$ & 1.5 & \phantom{0}1 & --- & 1500 \\
M3 & \phantom{0}3 & $-1.0$ & 1.5 & 20 & 0.5 & 1500 \\
M4 & 10 & $-3.0$ & 1.5 & 20 & 0.5 & 2300 \\
M5 & 10 & $-3.0$ & 1.5 & 50 & 0.5 & 2700 \\
    \bottomrule
    \tabnotetext{a}{Mass loss of the wind from the stellar companion}
    \tabnotetext{b}{Position on the $xz$-plane of the stellar wind source}
    \tabnotetext{c}{Asymmetry parameters of the wind}
\tabnotetext{d}{Time at which the expanding shell starts to leave
the computational grid (the frames of figures 2-6 correspond to these
times)}
  \end{tabular}
\end{table}

The source of the wind is located at a position $(x_w,0,z_w)$ away
from the origin of the coordinate system (which is at the center
of the computational domain, see Figure \ref{fig1} and \S 2.1). The
values of $x_w$ and $z_w$ chosen for the computed models are given
in Table 1. The whole of the computational domain (with the exception
of the region occupied by the ``bipolar pulse'', see \S 2.3) is
initialized with the wind solution, with the density and velocity
given by Equations (\ref{rho}-\ref{f}). Also, the wind solution is
imposed at all times within a sphere of radius $r_w={2.5\times 10^{16}}$~cm,
centered on the $(x_w,0,z_w)$ source position (see Figure \ref{fig1}).

We impose a uniform, $T_w=10^3$~K within the sphere of radius
$r_w$, and a temperature
\begin{equation}
T=T_w\,\left({r_w\over r}\right)^{2(\gamma-1)}\,,
\label{tw}
\end{equation}
outside of this sphere, where
$T_w$ and $r_w$ are constants defined above, $\gamma=5/3$ is the
specific heat ratio and $r$ is the spherical radius measured from
the position of the wind source. This temperature stratification
(corresponding to a constant velocity, adiabatic wind) is of course
only used in the $r>r_w$ region as an initial condition. Finally,
the ejected wind is assumed to have fully neutral H, and to have a small,
seed ionization fraction coming from singly ionized C.

\subsection{The bipolar pulse}

As an initial condition, we impose a ``bipolar pulse'' of material
ejected from a source located at the origin $(x,y,z)=(0,0,0)$
of the coordinate system (see Figure \ref{fig1}). This bipolar ``pulse'' is
imposed as an initial condition within a sphere of radius $r_c$
(centered on the origin, see Figure \ref{fig1}), and has a ``hubble-like velocity law'',
radially directed velocity of the form
\begin{equation}
v=v_c\,\left({R\over r_c}\right)\,,
\label{vc}
\end{equation}
where $R$ is the spherical radius (measured from the position of
the bipolar outflow source), and a density
\begin{equation}
\rho=\rho_c\,g(\eta)\,,
\label{rhoc}
\end{equation}
where
\begin{equation}
\eta={r_c\sin \theta_c \over R\sin \theta},\,
\label{eta}
\end{equation}
and
\begin{equation}
g(\eta)=\min[1,\eta]\,.
\label{g}
\end{equation}
Equations (\ref{rhoc}-\ref{g}) give a cylindrically stratified
density distribution, with a core of cylindrical radius
$r_c\sin \theta_c$ of constant density $\rho_c$, and
densities that fall monotonically for larger cylindrical radii.
This kind of density stratification for the jet-like flow
is inspired in the asymptotic MHD wind solution of Shu et al.
(1995).

In all of our models we have considered $r_c={5\times 10^{16}}$~cm,
$n_c=\rho_c/(1.3m_H)=10^4$~cm$^{-3}$ (where $m_H$ is the mass of the
hydrogen atom) for the central density of the ``collimated pulse''
(equation \ref{rhoc}), $\theta_c=5^\circ$ (equation \ref{eta})
and $v_c=200$~km~s$^{-1}$.
For the ``pulse'' material (within the sphere of radius $r_c$,
see above and figure \ref{fig1}) we have assumed a uniform temperature
of 100~K, and that the gas is neutral (except for a
seed ionization fraction coming from singly ionized C).

The bipolar ``pulse'' is then ``released'' at $t=0$ (i.~e.,
the material within the sphere of radius $r_c$ is allowed
to evolve freely as a function of time), and the
gasdynamic equations are integrated in time to follow the
interaction of the ``pulse'' with the stellar wind (described in
\S 2.2). The results of the numerical integrations are
described in the following section.

\begin{figure}[!t]
  \includegraphics[width=\columnwidth]{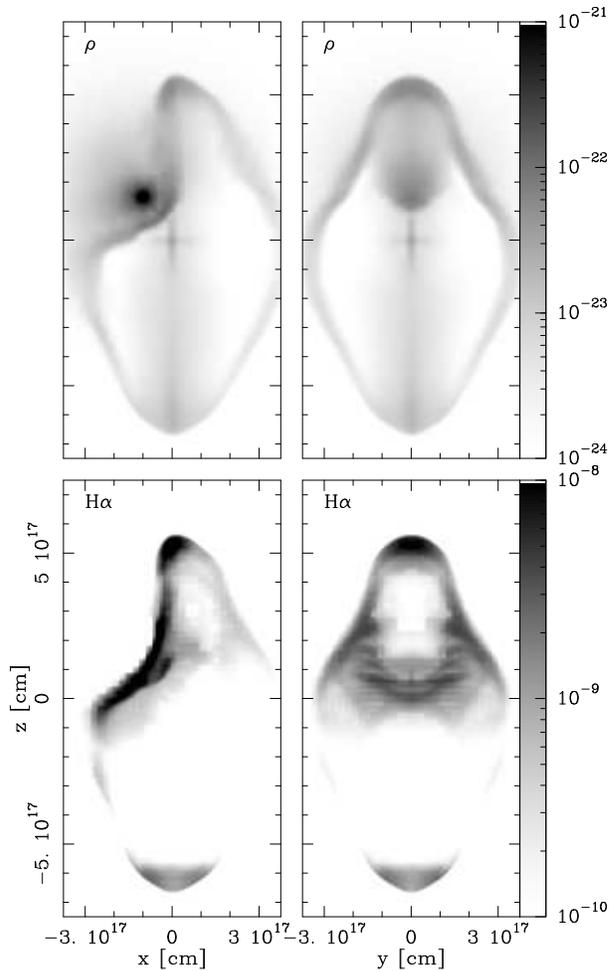}
  \caption{Results obtained from model M1 for a $t_f=1100$~yr
(see the text and table 1). The top frames show the density stratifications
on the $xz$- (left) and $yz$-plane (right) with a logarithmic greyscale
(given in g~cm$^{-3}$ by the bar on the right). The bottom frames
show the H$\alpha$ maps obtained by integrating the H$\alpha$ emission
coefficient along the $y$- (left) and $x$-axis (right) with a logarithmic
greyscale (given in erg~cm$^{-2}$~s$^{-1}$~sterad$^{-1}$ by the bar on
the right).}
  \label{fig2}
\end{figure}

\begin{figure}[!t]
  \includegraphics[width=\columnwidth]{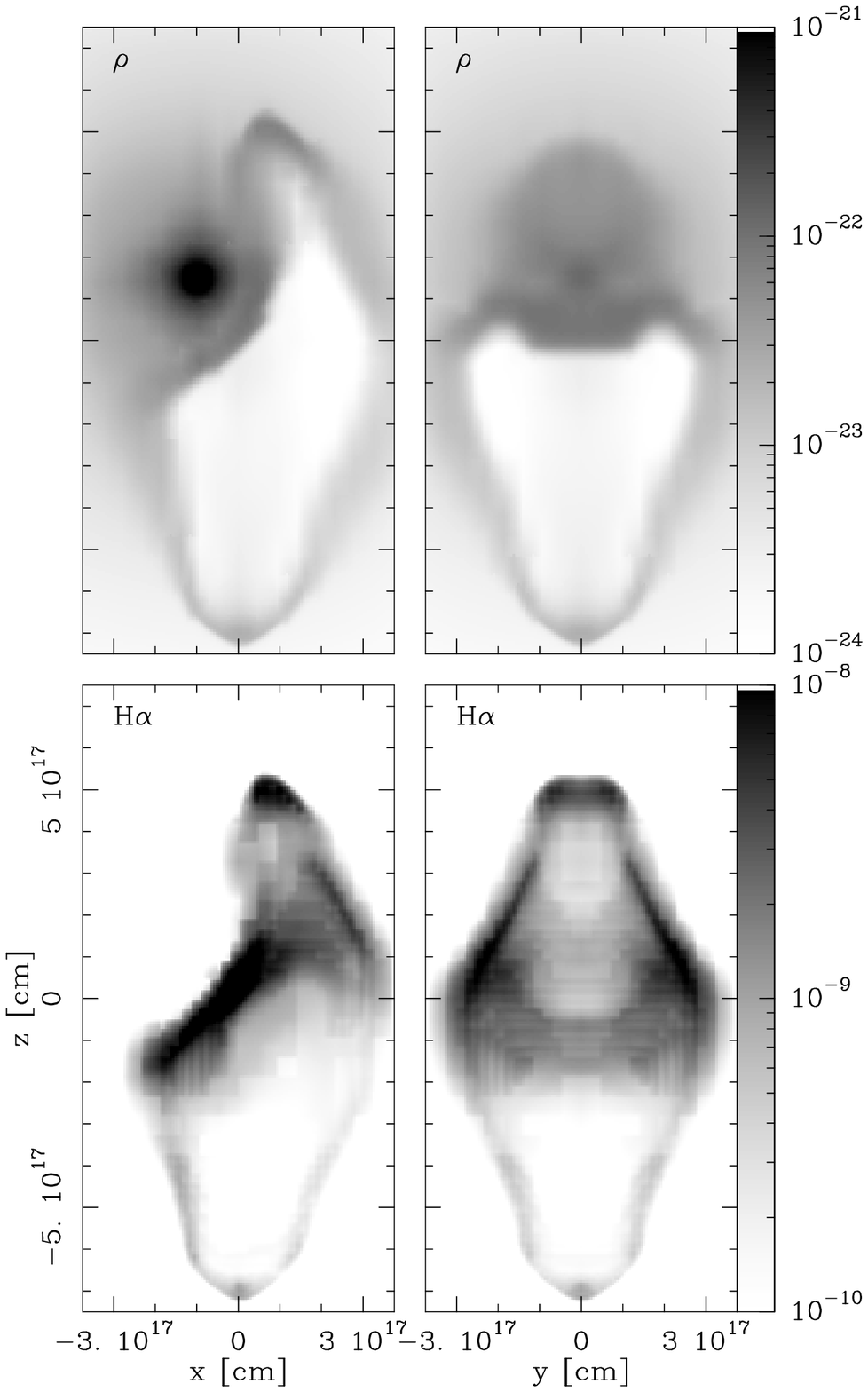}
  \caption{Results obtained from model M2 for a $t_f=1500$~yr
(see the text and table 1). A description of the four frames
is given in the caption of Figure 1.}
  \label{fig3}
\end{figure}

\begin{figure}[!t]
  \includegraphics[width=\columnwidth]{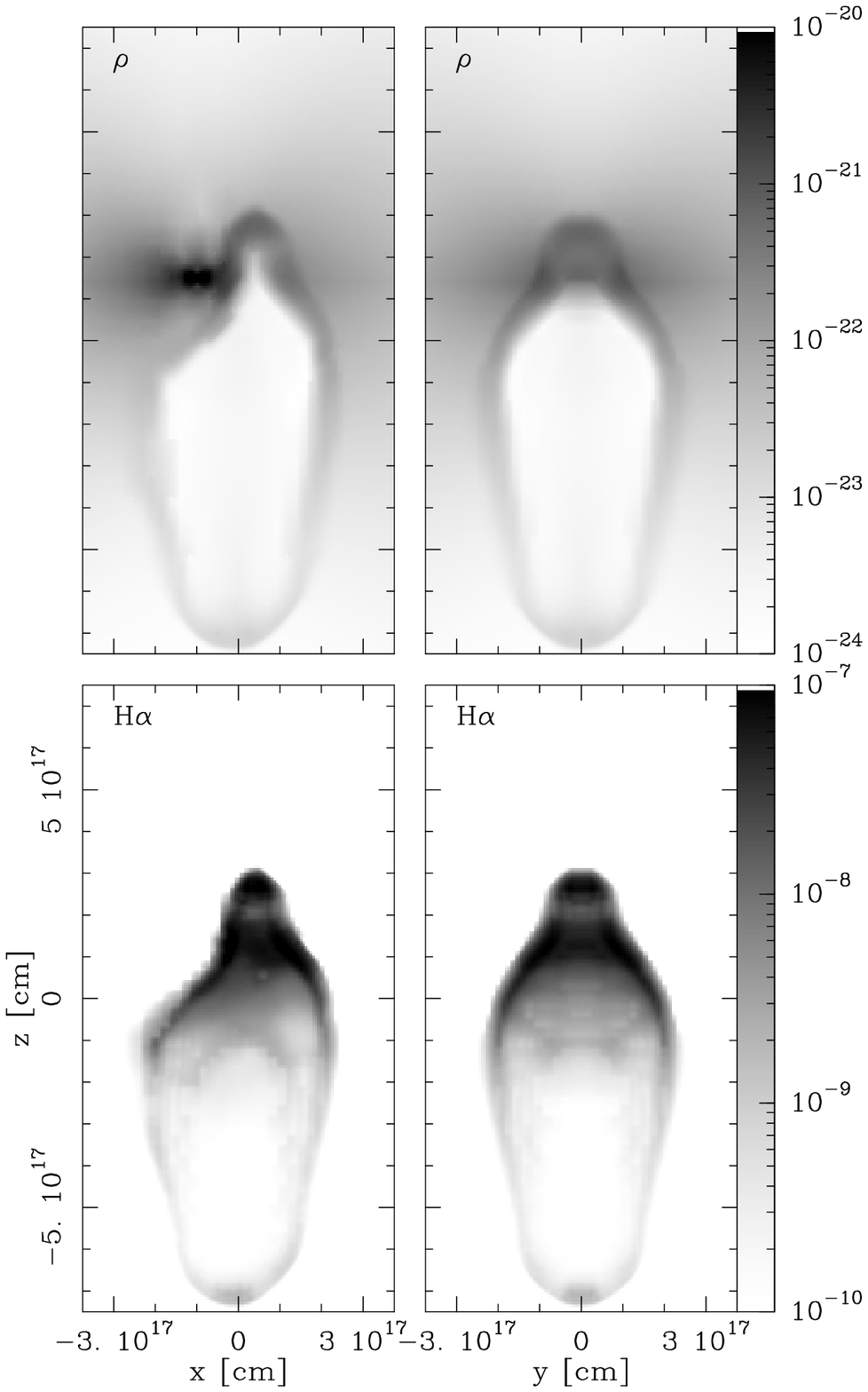}
  \caption{Results obtained from model M3 for a $t_f=1500$~yr
(see the text and table 1). A description of the four frames
is given in the caption of Figure 1.}
  \label{fig4}
\end{figure}

\begin{figure}[!t]
  \includegraphics[width=\columnwidth]{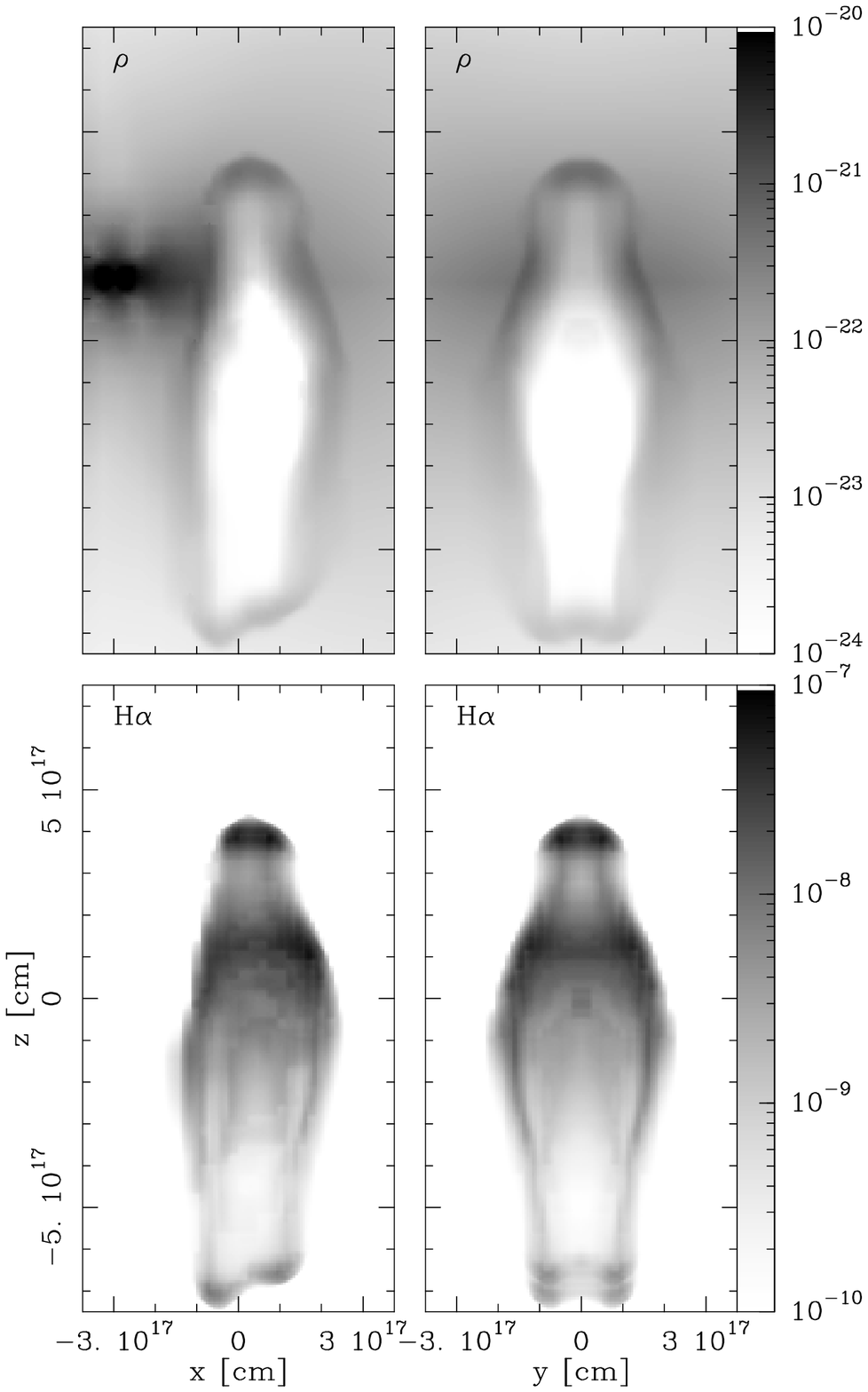}
  \caption{Results obtained from model M4 for a $t_f=2300$~yr
(see the text and table 1). A description of the four frames
is given in the caption of Figure 1.}
  \label{fig5}
\end{figure}

\begin{figure}[!t]
  \includegraphics[width=\columnwidth]{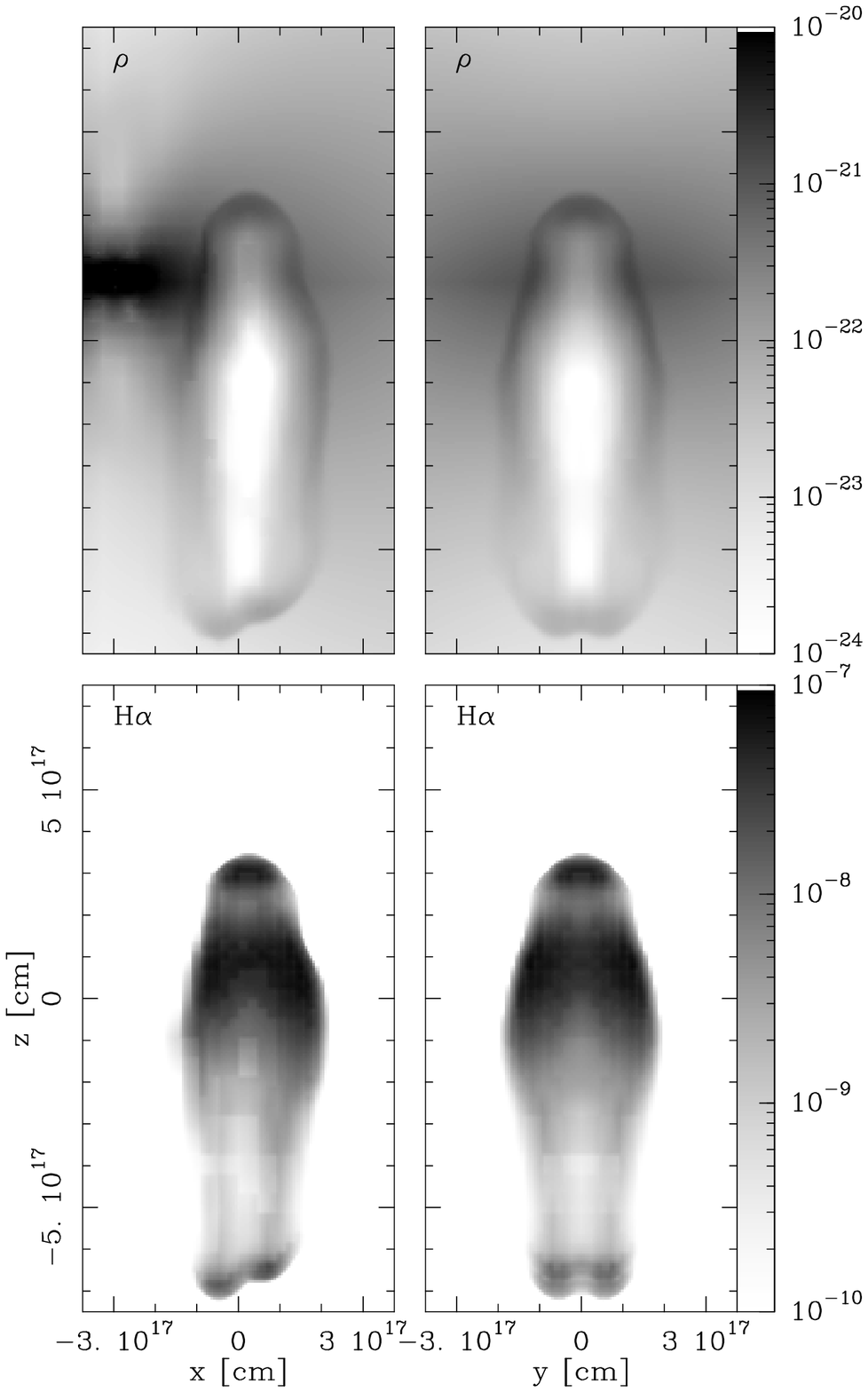}
  \caption{Results obtained from model M5 for a $t_f=2700$~yr
(see the text and table 1). A description of the four frames
is given in the caption of Figure 1.}
  \label{fig6}
\end{figure}

\section{Model results}

We have then computed 5 models (models M1-M5), with identical
``bipolar pulses'' (see \S 2.3) interacting with winds
of different parameters (coming from the binary companion
of the source of the bipolar pulse, see Figure \ref{fig1}). We have
tried both isotropic winds (models M1 and M2) and latitude
dependent winds (models M3-M5, see table 1), as described
in \S 2.2.

In Figures \ref{fig2}-\ref{fig6}, we show some of the results obtained from models
M1-M5 for the times $t_f$ (see Table 1)
at which the shell which is pushed out by
the bipolar ``pulse'' reaches one of the boundaries of the
computational domain. These Figures show the density stratifications
on the $xz$-plane (which includes the positions of both the
wind and the ``bipolar pulse'' sources, see Figure \ref{fig1}) and on
the $yz$-plane. These Figures also show the H$\alpha$ maps
obtained by integrating the H$\alpha$ emission coefficient
(including the radiative recombination cascade and the
$n=1\to 3$ collisional excitations) along the $y$- and
$x$-axes.

The models with spherically symmetric winds (models M1 and M2,
Table 1) produce highly asymmetric
$xz$-plane density stratifications and $xz$-H$\alpha$ emission
maps (left hand side frames of Figures \ref{fig2} and \ref{fig3}). The upper
lobe of the shell (which is pushed out by the bipolar pulse)
shows a major ``indentation'' formed by the interaction
with the dense region at the base of the stellar wind.
The $yz$-plane cuts and H$\alpha$ maps (right hand side
frames of Figures \ref{fig2} and \ref{fig3}) show an upper lobe which is
narrower and shorter than the bottom lobe, with a larger
asymmetry for the model with larger mass loss rate (model M2, see
table 1).

In model M3, we have a wind mass loss
rate intermediate between the ones of models M1 and M2 (see Table 1),
but we have considered a latitude-dependent wind (with
$\xi=20$ and $p=0.5$, see equation \ref{f}). In model M3 (see Figure \ref{fig4}),
the bottom lobe of the outflow is somewhat narrower, but otherwise similar
to the corresponding lobes of models M1 and M2 (see Figures \ref{fig2} and \ref{fig3}).
The upper lobe of model M3, however, is strongly reduced in size,
leading to a large asymmetry between the top and bottom lobes of
the expanding shell.

In models M4 and M5, we have moved the stellar wind source away from
the axis of the ``bipolar pulse'' ejection (see Table 1 and Figures
\ref{fig5} and \ref{fig6}). These models have a high, ${\dot M}_w=10^{-5}$M$_\odot$yr$^{-1}$
mass loss rate and two different degrees of asymmetry ($\xi=20$ and 50
for models M4 and M5, respectively). These models produce
top-to-bottom lobe asymmetries which are qualitatively similar to the ones
obtained from model M3 (see above and Figure \ref{fig4}).

As can be seen from the left hand side plots of Figures \ref{fig5} and \ref{fig6},
models M4 and M5 have the interesting property that the
side-to-side asymmetries of the top lobe (when seen in the $xz$-plane
density cut or H$\alpha$ map) are much less important than the
ones found in models M1-M3. This result indicates that a wind
with a strong latitude dependence from a source placed well away
from the axis of the ``bipolar pulse'' produces a strong
asymmetry between the two lobes of the expanding shell, but
does not introduce strong deviations from the initial axisymmetry
of the bipolar pulse.

Models M4 and M5 show a double peak structure at the bottom lobe 
(see Figures \ref{fig5} and \ref{fig6}), which are reminiscent of the results 
obtained from numerical simulations of stellar jets for a high 
jet to environment density ratio (e.g.,  Raga 1988;  Downes 
\& Cabrit 2003). 

\section{Discussion}

The results of our 3D numerical simulations of the interaction between
a bipolar ``pulse'' and a continous wind ejected by a companion
star indicate that the overall 
morphology (particularly, the strong differences between the 
sizes and velocities of the two lobes) of the PPN OH231.8+4.2 can be
modeled as the result of such an
interaction.   

We have adopted a bipolar ``pulse'' (described by equations 
\ref{vc}, \ref{rhoc}, \ref{eta}, and \ref{g}) 
instead of a continous jet, as an outflow of this kind appears
to be implied by the observed Hubble-like velocity law (see \S 1).
Also, we have taken the parameters
($\rho_c$, $v_c$, and $\theta_c$),
 of the dense, high-velocity jet (see \S 2.3) 
that would be necessary for explaining the CO observations of OH 231.8+4.2
of Alcolea et al. (2001). 

The parameters describing the wind from the companion star (i.e.,
QX Pup in the case of 
OH 231.8+4.2) are the wind velocity ($v_w$) and the mass loss rate ($\dot M_w$). 
We have adopted a value of 40 km s$^{-1}$ for the wind velocity, which corresponds 
to the escape velocity from QX Pup (Alcolea 2004) and mass loss rate values   
characteristic of Mira variables (with values ranging from 10$^{-6}$ to 10$^{-5}$ 
$M_\odot$ yr$^{-1}$; see Table 1). 
We have considered a stellar wind with a latitude dependent density, 
with a density enhancement at the equator. The equator is perpendicular to the 
``bipolar pulse'', as suggested by the spatial distribution of the SiO 
maser associated with QX Pup (S\'anchez Contreras et al. 2002).  

With these parameters we have computed several numerical simulations, 
from which we have obtained the H$\alpha$ emission maps, that can be compared with the 
H$\alpha$ images of OH 231.8+4.2 (S\'anchez Contreras et al. 2000a; Bujarrabal et al. 2002).

The numerical simulations reproduce qualitatively well several characteristics of the 
H$\alpha$ image of OH 231.8+4.2. The observed and predicted H$\alpha$
maps have bipolar structures with strong lobe-to-lobe asymmetries.
We should point out that 
the hourglass morphology of OH 231.8+4.2 is not reproduced by these models. This 
discrepancy can be solved by including a dense disk (or torus) around the central source 
(as has been shown numerically by several authors; see, e.g., Frank \& Mellema 1996).
However, we have not included such a torus in the present simulations.
  
All of our models show an upper lobe which is narrower and shorter 
than the bottom lobe, but  
the large asymmetry between the Northern and Southern lobes of OH 231.8+4.2 is only    
reproduced by the models with a latitude dependent wind (M3 - M5). Figures 
\ref{fig4}, \ref{fig5} and \ref{fig6} illustrate that 
the upper lobe is strongly reduced in size in the latitude-dependent
wind models, and that the ratio of $\sim 2$ between the sizes
 of the two lobes is in qualitative agreement with the
observations of OH 231.8+4.2 (Bujarrabal et al. 2002).
However, the width of the Southern lobe is better predicted by 
the models with spherically symmetric winds (models M1 and M2). 

\begin{figure}[!t]
  \includegraphics[width=\columnwidth]{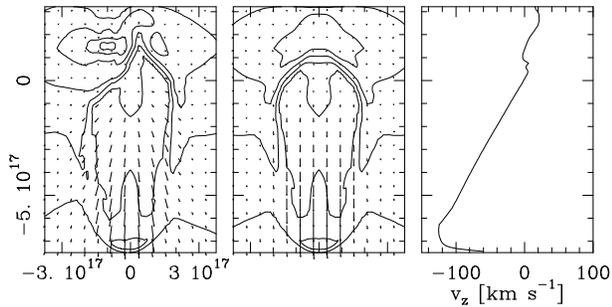}
  \caption{Results obtained for model M3 for a $t_f=1500$~yr
(see the text and table 1). The left and middle frames show the velocity field on the 
{\it xz}-plane (left) and {\it yz}-plane (middle). The velocity maps (arrows) are 
superposed on the  density stratification, with linearly spaced isolines with steps of 5 x 10$^{-22}$ 
g cm$^{-3}$. The right frame shows the axial velocity along the symmetry axis.}
  \label{fig7}
\end{figure}

To illustrate the predicted velocity field, we present the 
velocity field of model M3 on the {\it xz}-plane and {\it yz}-plane 
(see Figure \ref{fig7}). For a comparison with the 
observed properties we also present the axial 
velocity along the symmetry axis (see Figure\ref{fig7}). The predicted kinematics 
reproduces the strong velocity gradient observed along the 
nebular axis (Alcolea et al. 2001; S\'anchez Contreras et al. 2000). 
 However, we find that the model prediction gives velocities lower than the 
observed ones (by a factor of 3 to 4). In the top lobe, 
the model produces a too sharp drop in the velocity. 
In order to obtain a better agreement with the observations, we would need to 
 have a model with a higher velocity bipolar pulse. 

We have presented a number of 3D simulations of a bipolar
pulse interacting with a continuos 
wind ejected by a close companion star. In our simulations,
we have used a dense, AGB wind 
with mass loss rates typical of AGB stars.
We have shown that the strong asymmetries between both lobes of 
this nebula can indeed be explained in terms of a the proposed
bipolar pulse/continous wind interaction model (the companion
star ejecting the wind has to be located within the Northern lobe).
We find that the observed lobe-to-lobe asymmetry can be explained
without invoking an intrinsic asymmetry in the ejection of the
jet and the counterjet.

\medskip

\acknowledgements
The work of ACR and was supported by the CONACyT grants 36572-E
and 41320 and the DGAPA (UNAM) grant IN~112602.
The work of ARi was supported by the Spanish MCyT grant AYA2002-00205. 
We acknowledge an anonymous referee for her/his helpful comments. 
\linebreak

\vspace*{\baselineskip}


\end{document}